\documentclass[aps,prb,reprint,superscriptaddress]{revtex4-1}
\usepackage{mathptmx,siunitx,graphicx,amssymb,natmove}
\usepackage[version=3]{mhchem}

\def\df{d_\text{f}}
\def\Ef{E_\text{f}}

\def\sigf{\sigma_\text{f}}
\def\thes{\theta_\text{s}}
\def\Le{L_\text{e}}

\DeclareSIUnit{\tex}{tex}
\DeclareSIUnit{\sq}{sq}

\begin{document}
\title{Assembly-Dependent Interfacial Property of Carbon Nanotube Fibers with Epoxy and Its Enhancement via Generalized
Surface Sizing}

\author{Chaoshuai Lei}
\affiliation{Key Laboratory of Aerospace Advanced Materials and Performance, School of Materials Science and
Engineering, Beihang University, Xueyuan Road 37, Beijing 100191, China}
\author{Jingna Zhao}
\author{Jingyun Zou}
\author{Chunyang Jiang}
\affiliation{Key Laboratory of Nano-Devices and Applications, Suzhou Institute of Nano-Tech and Nano-Bionics, Chinese
Academy of Sciences, Ruoshui Road 398, Suzhou 215123, China}
\author{Min Li}
\affiliation{Key Laboratory of Aerospace Advanced Materials and Performance, School of Materials Science and
Engineering, Beihang University, Xueyuan Road 37, Beijing 100191, China}
\author{Xiaohua Zhang}
\email{xhzhang2009@sinano.ac.cn}
\affiliation{Key Laboratory of Nano-Devices and Applications, Suzhou Institute of Nano-Tech and Nano-Bionics, Chinese
Academy of Sciences, Ruoshui Road 398, Suzhou 215123, China}
\author{Zuoguang Zhang}
\affiliation{Key Laboratory of Aerospace Advanced Materials and Performance, School of Materials Science and
Engineering, Beihang University, Xueyuan Road 37, Beijing 100191, China}
\author{Qingwen Li}
\affiliation{Key Laboratory of Nano-Devices and Applications, Suzhou Institute of Nano-Tech and Nano-Bionics, Chinese
Academy of Sciences, Ruoshui Road 398, Suzhou 215123, China}

\begin{abstract}
In building up composite structures using carbon nanotube (CNT) fibers, the fiber-to-matrix interfacial shear strength
(IFSS) is one of the most important issues. Originating from the assembly characteristics of CNT fiber, the IFSS
strongly depends on the fiber's twisting level and densification level. Furthermore, there are rich ways to modify the
fiber surface and thus enhance the IFSS, including the physical and chemical modification on fiber surface, the
infiltration of matrix resin into CNT fiber, and the introduction of silane coupling agent. A new feature differing from
carbon fibers is that all these treatments either change the fiber's surface structure or form a
several-hundreds-nm-thick interphase inside rather than around the fiber. These ``generalized'' treatments obviously
extend the common concept of surface sizing and can be used for various forms of CNT assembly structures.
\end{abstract}

\maketitle

\section{Introduction}

Carbon nanotubes (CNTs) have gained extensive attention since their discovery owing to their extraordinary mechanical
and electrical properties \cite{ruoff.rs:2003}. As an important progress, macroscopic and continuous fibers containing
millions of individual CNTs have been reported and studied in the past decade \cite{zhang.xh:2012, lu.wb:2012,
miao.mh:2013}. There are three main strategies for spinning CNT fibers, namely, the coagulation-based ``wet spinning''
\cite{vigolo.b:2000, dalton.ab:2003, ericson.lm:2004}, the ``direct spinning'' from a CNT aerogel \cite{zhu.hw:2002,
li.yl:2004, koziol.k:2007} or similarly from a pre-formed CNT film \cite{ma.wj:2009}, and the ``array spinning'' based
on vertically aligned CNT arrays (forests) \cite{jiang.kl:2002, zhang.m:2004, zhang.xb:2006, li.qw:2006}. Owing to the
long CNT length, high CNT alignment, and availability of various post-spin strengthening treatments, the array-spun CNT
fibers have demonstrated the best mechanical performances to date \cite{zhao.jn:2010, fang.c:2010, jia.jj:2011,
li.s:2012, meng.fc:2014}.

From the structural comparison between CNT fiber and carbon fiber, it is believed that the interfacial property between
CNT fiber and matrix is much better than that in carbon fiber-reinforced polymer. This is because the assembly
characteristics of CNT fiber can introduce rich interfacial contacts. Unfortunately, the interfacial shear strength
(IFSS) between CNT fiber and epoxy was just about 14.4--20.2 MPa \cite{deng.f:2011, zu.m:2012}. The low IFSS was owing
to the shortcomings of characterization technique and the unsized fiber surface. For the former issue, there are two
widely used micro-mechanical techniques to measure the IFSS, namely the single-fiber fragmentation and microdroplet test
\cite{herrera-franco.pj:1992, pitkethly.mj:1993} The single-fiber fragmentation often meets a problem of strain mismatch
between the fiber and matrix, as the matrix should have an ultimate tensile strain much larger than that of the fiber to
avoid the fracture-induced failure \cite{herrera-franco.pj:1992, pitkethly.mj:1993, drzal.lt:1993, tripathi.d:1996}.
Therefore the microdroplet test is more preferred for CNT fibers whose ultimate strain is large (ranging from
$\sim$3--5\% to over 10\%), unless a matrix with a quite high fracture strain is used \cite{liu.yn:2013}. Such test can
also be easily used to test the interfacial properties with different polymer matrices and to analyze the debonding
phenomenon at the interface \cite{zu.m:2012}. On the other hand, although the sizing treatment is important in the
manufacturing process of carbon fibers, it is unfortunately seldom performed for CNT fibers as the interfacial binding
energy can be maximized due to the rich surface areas from the assembly structure.

This paper is intended to uncover the assembly-dependent interfacial properties of CNT fiber by using microdroplet
tests. Owing to the assembly characteristics, a variety of approaches can be used to form enhanced bonding between CNT
fiber and matrix, such as physical and chemical modification on fiber surface, resin infiltration into CNT fiber, and
the usage of silane coupling agent. As compared to the common sizing treatment that usually refers to the interphase
formation between the fiber and the matrix in a composite, the treatments here form a several-hundreds-nm-thick
interphase inside CNT fibers by impregnating the matrix resin or coupling agent, and thus extend the concept of surface
sizing. These generalized sizing treatments increase the CNT-to-matrix contact area and thus the IFSS. For example, for
the dry-spun and liquid-densified CNT fibers, the IFSS with epoxy was 30.9--44.4 MPa and 43.5--53.1 MPa, respectively.
After being sized with silane coupling agent, the IFSS was improved up to 58.8--68.2 MPa. Furthermore, as CNTs can form
various macroscale assemblies \cite{liu.lq:2011}, the generalized concept of surface sizing can serve as a new efficient
way to build high performance structural composite materials.

\section{Results and Discussion}

\subsection{Microdroplet test}

Different from carbon fiber, CNT fiber is an assembly of millions of individual nanotubes and thus has a totally
different surface morphology. The assembly characteristics makes it a paradox to calculate the real surface area, quite
like that the coastline length depends on the method used to measure it. Therefore the IFSS between CNT fiber and matrix
is usually the ``effective'' strength by dividing the maximum force at the onset of microdroplet debonding along the
fiber ($F_d$) by the fiber's circumference \cite{gaur.u:1989},
\begin{equation}
\text{IFSS} = \frac{F_d}{\pi \df \Le},
\end{equation}
where $\df$ and $\Le$ are the fiber's diameter and embedded fiber length by the microdroplet, respectively. Another
significant difference between CNT and carbon fibers is the helical assembly of CNT fiber, usually described by the
``surface'' twist angle $\thes$ \cite{zhao.jn:2010}. This angle results in stress loss along the fiber axis according to
the factor of $\cos\thes$. Further, as discussed above, the larger ultimate strain of CNT fiber requires a microdroplet
test rather than a single-fiber fragmentation to quantitatively measure the IFSS of CNT fiber with different matrices.

\begin{figure}[!t]
\centering
\includegraphics[width=.48\textwidth]{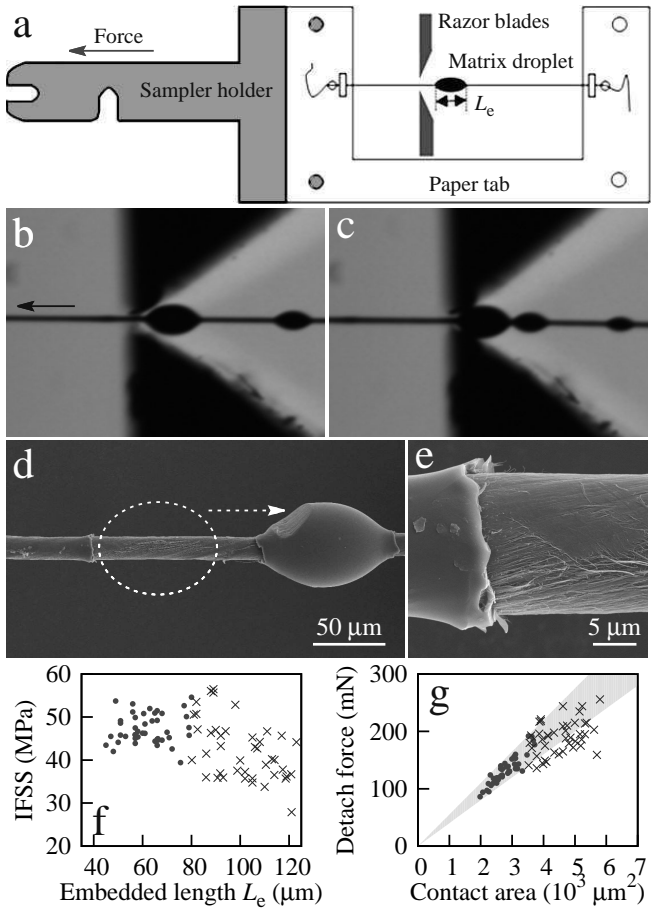}
\caption{\label{fig.microdroplet} Microdroplet test to evaluate the IFSS between a CNT fiber and epoxy matrix. (a)
Schematic of test set-up. (b,c) Optical microscopy images before and after a microdroplet test. (d,e) SEM images showing
that the matrix droplet debonded as a whole from the surface of an unsized CNT fiber. (f) IFSS as a function of $\Le$
for unsized CNT fibers. The results for $\Le<80$ and $>80$ were plotted with different symbols. (g) The linear
dependence between the detach force and contact area was observed for $\Le<80$ while became deviated otherwise.}
\end{figure}

Figure \ref{fig.microdroplet}a shows schematically the microdroplet test set-up, which is specially designed for the
Equipment for Evaluation of Fiber/Resin Composite Interface Properties (model HM410, Tohei Sangyo Co., Ltd., Tokyo,
Japan). Microdroplets were formed along a CNT fiber by curing the epoxy resins which were brushed on the fiber surface,
and the droplet sizes varied from below 20 to over 150 \si{\micro\m}. To find out the optimal droplet size, IFSS
measurements were performed for the droplets with $\Le$ ranging from 30 and 130 \si{\micro\m}. Figure
\ref{fig.microdroplet}b and c show the optical microscopy images before and after a microdroplet test, where a droplet
with $\Le \approx 80$ \si{\micro\m} slid from its original position to the left side of the neighboring droplet, and
their corresponding scanning electron microscopy (SEM) images are shown in Figure \ref{fig.microdroplet}d and e.

The optimal $\Le$ was 40--80 \si{\micro\m} according to Figure \ref{fig.microdroplet}f. When $\Le \ll 40$ \si{\micro\m},
the size effect arose from the enhanced system error for $\Le$ and the droplet can be easily damaged by the razor blades
before detaching from the fiber surface. On the other hand, for $\Le \gg 80$ \si{\micro\m}, the force to debond the
fiber-to-matrix interface was high enough to introduce Poisson effect (transverse shrinkage), resulting in a new
mechanism to debond the interface. This means that the detach force became slight smaller from the linear dependence on
the interfacial contact area (Figure \ref{fig.microdroplet}g). Furthermore, when $\Le$ became much larger, the debonding
force could be larger enough to break the CNT fiber. For example, for a CNT fiber with $\df\approx15$ \si{\micro\m} and
a tensile strength of $\sim$1.2--1.5 GPa, the fracture force was about 212--265 mN. This value corresponded to the upper
limit of $\sim$100--125 \si{\micro\m} for $\Le$ (when IFSS $\approx45$ MPa).

\subsection{Effect of CNT assembly structure}

The assembly characteristics of CNT fiber rather than the \ce{C-C} sp$^2$ lattice structure plays the key role in
determining the fiber-to-matrix interactions. According to our previous study, there are two important spinning
parameters related to the tensile properties, namely the fiber's diameter $\df$ and the twist angle $\thes$
\cite{zhao.jn:2010}. Furthermore, the CNT packing density can be increased by using liquid densification during the
spinning process \cite{li.s:2012}. For simplicity, here $\df$ was always controlled to be 14--16 \si{\micro\m}.
Therefore the comparison was performed on the twisting level and packing density.

The first comparison was performed on the tensile properties, which has also been systematically studied previously
\cite{zhao.jn:2010, fang.sl:2010}. In the present study, when $\thes\approx10$, 17, 23, and 30\si{\degree}, the dry-spun
CNT fibers exhibited tensile strengths $\sigf=753$, 868, 1032, and 980 MPa, and elastic moduli $\Ef=32.5$, 36.3, 43.5,
and 27.1 GPa, respectively. If ethylene glycol (EG) was used to assist the spinning\cite{li.s:2012}, both $\sigf$ and
$\Ef$ were remarkably improved. At the optimal $\thes\approx23$\si{\degree}, the EG-densified CNT fibers exhibited
$\sigf=1580$ MPa and $\Ef=54.5$ GPa.

\begin{figure*}[!t]
\centering
\includegraphics[width=.70\textwidth]{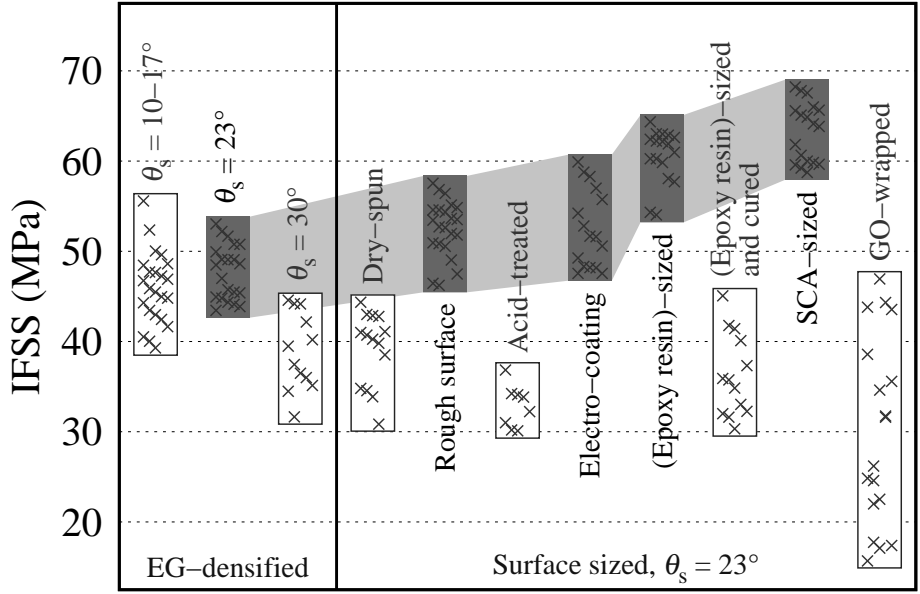}
\caption{\label{fig.IFSS} Effect of surface morphology and sizing treatment on the fiber-to-epoxy IFSS. Two comparisons
are provided for the effect of twisting angle and for the different surface treatments, respectively. For a better eye
view, those enhanced IFSS was highlighted by using the gray background.}
\end{figure*}

\begin{figure}[!t]
\centering
\includegraphics[width=.48\textwidth]{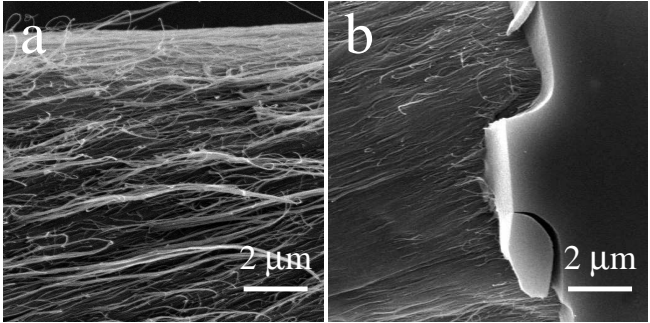}
\caption{\label{fig.sem} SEM images where the epoxy droplets had been removed from the fiber surface, for dry-spun (a)
and EG-densified (b) CNT fibers whose twist angles were $\approx23$\si{\degree}.}
\end{figure}

The different packing density also affected the fiber-to-matrix IFSS, see the results for dry-spun and EG-densified CNT
fibers shown in Figure \ref{fig.IFSS}. The averaged, minimum, and maximum values of IFSS, and the corresponding standard
variation are also provided in Table \ref{tab.IFSS}. The low densification level of the dry-spun fibers strongly limited
the inter-tube interactions, and thus the CNTs was easy to be peeled off from the fiber. Figure \ref{fig.sem}a shows an
SEM image for a tested dry-spun fiber ($\thes\approx23$\si{\degree}) where the peeled CNTs were observed after removing
the epoxy droplet. The measured IFSS was 30.9--44.4 MPa (average 39.2 MPa). For the EG-densified CNTs, the improved
densification increased the inter-tube interactions and thus the fiber's mechanical performance. When the droplet was
removed, the peeling-off phenomenon nearly disappeared and the sliding of the droplet did not damage the fiber surface,
see Figure \ref{fig.sem}b and Figure \ref{fig.microdroplet}d,e where both fibers were EG-densified. This indicated that
the IFSS should be higher, in agreement with the measurement for $\thes\approx23$\si{\degree} (43.5--53.1 MPa, average
47.7 MPa).

\begin{table}
\caption{\label{tab.IFSS} The averaged, minimum, and maximum values of IFSS measured for different CNT fibers and the
corresponding standard deviation, in unit of MPa.}
\centering
\begin{tabular}{ccccc}
\hline\hline
Fiber type    &      &         &         & Standard \\
           or & IFSS & Minimum & Maximum & deviation \\
surface treatment & (MPa) & (MPa) & (MPa) & (MPa) \\
\hline
\multicolumn{5}{l}{EG-densified CNT fibers}\\
$\thes\approx10$--17\si{\degree} & 46.0 & 39.3 & 55.6 & 3.88 \\
$\thes\approx23$\si{\degree}     & 47.7 & 43.5 & 53.1 & 2.95 \\
$\thes\approx30$\si{\degree}     & 38.9 & 31.7 & 44.6 & 4.12 \\
\hline
\multicolumn{5}{l}{Surface-sized CNT fibers}\\
Dry-spun            & 39.2 & 30.9 & 44.4 & 3.94 \\
Rough surface       & 51.5 & 46.3 & 57.6 & 3.09 \\
Acid-treated        & 32.8 & 30.1 & 36.8 & 2.20 \\
electro-coating     & 52.5 & 47.6 & 59.9 & 4.19 \\
(Epoxy resin)-sized & 60.6 & 54.0 & 64.4 & 2.90 \\
(Epoxy resin)-sized &      &      &      &      \\
and cured           & 36.3 & 30.3 & 45.1 & 4.43 \\
SCA-sized           & 63.3 & 58.8 & 68.2 & 3.20 \\
GO-wrapped          & 29.9 & 15.7 & 47.0 & 10.19 \\
\hline\hline
\end{tabular}
\end{table}

By focusing on the EG-densified fibers, a similar dependence on $\thes$ was also observed where the optimal
23\si{\degree} angle induced the strongest and most stable fiber-to-matrix interface. As shown in Figure \ref{fig.IFSS}
and Table \ref{tab.IFSS}, the IFSS for $\thes\approx10$--17\si{\degree} was 39.3--55.6 MPa (average 46.0 MPa), and above
30\si{\degree}, the IFSS decreased remarkably to 31.7--44.6 MPa (average 38.9 MPa). (The results for 10\si{\degree} and
17\si{\degree} were mixed together as the value differences were small.)

\subsection{Generalized surface sizing}

Surface sizing is a common treatment for carbon fibers to protect them from damage and to improve the interfacial
properties with polymer matrix. In this treatment, the sizing agent adheres to substrate fibers and can bond with the
surrounding polymer molecules. In order to improve the IFSS of CNT fibers, similar treatment can also be applied. Here
we show that, by considering the assembly characteristics, there are various and different ways to size CNT fibers,
building up a ``generalized'' sizing concept. The methods are (1) physical and chemical modification on fiber surface,
(2) resin infiltration into CNT fibers, and (3) sizing with silane coupling agent.

\subsubsection{Surface modulation without using sizing agents}

\begin{figure}[!t]
\centering
\includegraphics[width=.48\textwidth]{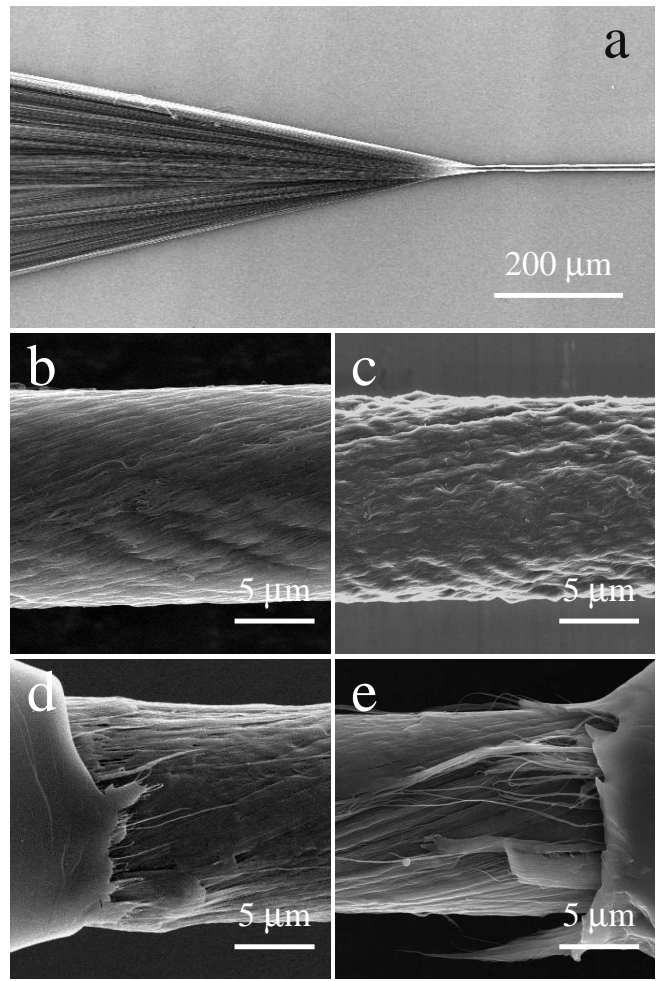}
\caption{\label{fig.rough} (a) A CNT sheet became triangular under twisting to form a fiber. (b,c) SEM images for two
EG-densified CNT fibers with smooth and rough fiber surfaces, respectively. (d) For an acid-treated fiber, the overall
contact between the fiber and droplet was weak as there was no damage like peeling-off on the fiber surface. (e) The
infiltration with dilute resin/acetone solution formed a several-hundreds-nm-thick interphase which could be observed
from the peeling-off phenomenon.}
\end{figure}

It is possible to increase the interfacial contact area by roughening the fiber surface. As discussed above, although
the dry-spun fibers had a rough surface due to the low densification, the peeling-off phenomenon restricted the IFSS to
be no more than 45 MPa. Thus it was necessary to roughen the surface of the EG-densified fibers. The rough and smooth
fiber surfaces were produced by adjusting the position to introduce the liquid infiltration. As reported previously
\cite{jia.jj:2011}, a CNT sheet is drawn out of a spinnable CNT array and then becomes triangular under twisting, see
Figure \ref{fig.rough}a. Generally, liquid infiltration is applied at the end of or slightly after the triangular zone.
As the fiber has been formed under twisting, the liquid just improves the packing density and smoothes the fiber surface
(Figure \ref{fig.rough}b). However, when the infiltration was applied before the triangular zone, the densification
would make the fiber formation earlier, resulting in a competition between the densification and twisting. As a result,
the fiber surface became much rougher, see Figure \ref{fig.rough}c. (Fortunately, the roughening just lowered the
tensile strength by less than 100 MPa.) As the rough surface increased the contact area and thus the interfacial binding
energy, the measured IFSS increased up to 46.3--57.6 MPa (Figure \ref{fig.IFSS}).

Chemical functionalization is another way to modulate the fiber surface, like an acid treatment \cite{meng.fc:20121}.
The surface functionalization (such as hydroxyl and carbonyl groups) might introduce covalent bonding between the fiber
and matrix and thus increase the IFSS. For example, the functionalization could improve the surface wettability and
affinity to the resin, and thus benefit the IFSS. The densified and roughened surface could also have the similar effect
as the rough EG-densified fibers did. Unfortunately, as shown in Figure \ref{fig.IFSS}, a remarkable decrease in IFSS
(only 30.1--36.8 MPa) was observed for the acid-treated CNT fibers. Thus we suspect that besides the strong covalent
bonding, a new and weak interface was formed in the composite structure. The reason can be analyzed as below. Due to the
functional groups on the fiber surface, there were also reactions between resin molecules with them during the curing
process. However, as it was difficult for the CNT to change its molecule position, the reaction should be finished by
the self-moving of resin molecules. As a result, the resins reacted with the CNT could not react sufficiently with other
resins, corresponding to a weak interaction in the composite structure. In other words, the covalent bonding resulted in
an epoxy-wrapped composite fiber, and there existed a weak interface between this composite fiber with the outer epoxy
droplet. Such analysis can be confirmed by SEM images. Figure \ref{fig.rough}d shows that there were still a certain
epoxy maintained around the fiber (corresponding to the epoxy-wrapped composite fiber), as the outer droplet had been
pulled off due to the weak interface between the new fiber and the droplet.

Besides these two ways to directly modify the fiber surface, we also tried an electro-coating method where an electric
current (5--6 mA) was conducted to pass through the fiber. Due to the electro-thermal coupling effect
\cite{meng.fc:2014}, the fiber temperature could be 200--250 \si{\celsius}, as high enough to cure epoxy resins. The
fiber temperature was higher than the temperature (150 \si{\celsius}) used to thermally cure the epoxy, as the heat
dissipated very quickly from the fiber to the environment. On the other hand, the fiber temperature was not higher than
the glass transition temperature of epoxy (236--287 \si{\celsius}, peak at 270.4 \si{\celsius}) and much lower than the
decomposition temperature of epoxy ($\sim$310--320 \si{\celsius}), and thus such treatment did no harm to the epoxy. As
the electro-thermal coupling caused in situ curing on the fiber surface, the final IFSS increased up to 47.6--59.9 MPa,
after a 60-min treatment, as shown in Figure \ref{fig.IFSS} and Table \ref{tab.IFSS}.

These treatments on CNT fibers can efficiently modify the fiber-to-matrix interaction, and can be considered as a
``sizing'' treatment without using any sizing agent, However, these ``gentle'' sizing treatments could only improved the
IFSS by 6--7 MPa; it was difficult to form a tightly bonded interface from the direct introduction of epoxy resins
around CNT fibers.

\subsubsection{Infiltrating sizing treatment}

\begin{figure}[!t]
\centering
\includegraphics[width=.48\textwidth]{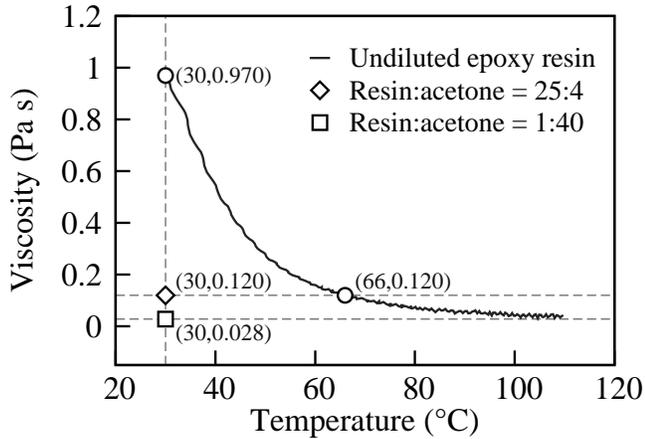}
\caption{\label{fig.viscosity} Viscosity for the raw and acetone-diluted resins.}
\end{figure}

In a previous study, epoxy resins were found to infiltrate into the fiber surface with a depth $\sim$1 \si{\micro\m} to
form an interphase, and thus to increase the contact area between CNT and epoxy. In the present study, however, as the
CNT fibers were highly densified with EG, the directly coated resins could hardly infiltrate into the fibers. In order
to introduce a certain depth of infiltration, the epoxy resins were diluted (wetted) with acetone (resin to acetone mass
ratio 25:4). The wetting also improved the rheological property of epoxy resin; its viscosity at room temperature
decreased from 0.97 to 0.12 \si{\Pa \s} (Figure \ref{fig.viscosity}). As acetone has high mobility and can quickly wet
CNT assemblies \cite{li.s:2012}, this treatment increased the IFSS. However, the improvement of about 6 MPa was not
high.

Based on the effect of acetone dilution, we developed a new sizing treatment. Before coating epoxy resins, we brushed
CNT fibers with much more diluted resin/acetone solutions (resin to acetone mass ratio 1:40) whose viscosity was only
0.028 \si{\Pa \s} (Figure \ref{fig.viscosity}). As more solvent could infiltrate into the fiber and thus take epoxy
resins to a certain depth, an interphase was formed to enhance the interfacial interaction. This treatment was totally
different from the common sizing concept where the sizing agent usually adheres to the fiber and forms a film around it.
Obviously, this is one advantage from the assembly characteristics of CNT fiber. After the wet resins (mass ratio 25:4)
was introduced on the fiber surface, the whole sample was ready to be cured.

The sizing with dilute solution increased significantly the IFSS, ranging from 54.0 to 64.4 MPa (average 60.6 MPa), see
Figure \ref{fig.IFSS}. The improvement (above the EG-densified fibers) was about 11--13 MPa, much higher than that from
the direct physical and chemical modification on fiber surface. The infiltration-induced interphase can be clearly
observed with SEM from the peeling-off (Figure \ref{fig.rough}e), and its thickness (corresponding to the depth of
polymer penetration) was just about 100--200 nm, much thinner than that ($\sim$1 \si{\micro\m}) observed in dry-spun
fibers \cite{zu.m:2012}.

If the sized fiber was cured before introducing the droplet, a weak fiber-to-matrix interface was measured on the
contrary (IFSS $\approx30.3$--45.1 MPa). This is not surprising as there was a new interface formed between epoxy due to
the two curing treatments, a situation quite similar to that in the acid-treated fibers.

\subsubsection{Sizing with coupling agents}

A coupling agent is a compound that can provide a chemical bond between two dissimilar materials. For example, silane
coupling agent (\ce{R(CH2)_nSiX3}, R and X being organofunctional and hydrolyzable groups, respectively) is most
commonly used in fiber surface treatment. Due to the assembly structure, the sizing treatment on CNT fibers using
coupling agents exhibited some new features. Typically, for (3-aminopropyl)triethoxysilane (APTES,
\ce{H2N(CH2)3Si(OCH2CH3)3}), the 3-aminopropyl group can chemically bonded with epoxy resin while the rich ethoxy groups
can increase the non-covalent interaction with CNT. The silanization process can also increase the bonding energy
between CNT and APTES once there is a certain functionalization on CNT \cite{kathi.j:2008}. Furthermore, the smaller
molecular weight (221.37 \si{\g\per\mol}) and lower viscosity ($<$0.01 \si{\Pa \s}) of APTES also allow very efficient
infiltration into CNT fibers. These advantages resulted in the highest IFSS of 58.8--68.2 MPa (average 63.3 MPa,
improvement over 15 MPa), see Figure \ref{fig.IFSS}. Due to the same densification level, the thickness of the APTES/CNT
interphase was also about 100--200 nm. Clearly, the infiltration-based sizing treatment inspires new strategies for
improving the interfacial interactions for various CNT assembly structures.

On the contrary, when non-penetrating polymers were used, the sizing treatment could just form a thin layer around the
fiber surface. For example, graphene oxide (GO) nanosheets can wrap a CNT fiber and act as a surface modifier
\cite{meng.fc:2012}. However, the IFSS between epoxy and GO-wrapped CNT fiber was quite low, ranging from 15.7 to 47.0
MPa (average 29.9 MPa), see Figure \ref{fig.IFSS}. As there was no polymer penetration into CNT fibers and the
functional groups of GO could covalently bonded with epoxy, the interface between CNT fiber and GO nanosheets became the
weakest part during the droplet test.

On the contrary, when non-penetrating polymers were used, the sizing treatment could just form a thin layer around the
fiber surface. For example, graphene oxide (GO) nanosheets can wrap a CNT fiber and act as a surface modifier
\cite{meng.fc:2012}. However, the IFSS between epoxy and GO-wrapped CNT fiber was quite low, ranging from 15.7 to 47.0
MPa (average 29.9 MPa), see Figure \ref{fig.IFSS}. As there was no polymer penetration into CNT fibers and the
functional groups of GO could covalently bonded with epoxy, the interface between CNT fiber and GO nanosheets became the
weakest part during the droplet test. Furthermore, as the size of GO was a few tens of \si{\micro\m}, larger than the
fiber's diameter, the wrapping level of GO might be different from fiber to fiber \cite{meng.fc:2012}. This is why a
large standard variation in IFSS was observed in the droplet tests (about 10.19 MPa, see Table \ref{tab.IFSS}).

\section{Conclusion}

Surface sizing is an important way to improve the interfacial interactions. Different from carbon fibers which are solid
in structure, the assembly characteristics of CNT fibers induces several new sizing treatments, including the physical
and chemical modification on fiber surface, the infiltration of matrix resin into CNT fiber, and the introduction of
silane coupling agent. These methods generalized or extended the common concept of surface sizing as they modified the
fiber structure from its surface to a certain depth (about 100--200 nm) rather than forming an additional interphase
around the fiber. By using the generalized treatments, the IFSS between CNT fiber and epoxy could be improved from
43.5--53.1 MPa to 58.8--68.2 MPa.

\section{Experimental Section}

The CNT fibers were spun by collecting the CNT sheet drawn out from vertically aligned CNT arrays under continuous
twisting \cite{zhao.jn:2010}. The CNTs were mainly double- to triple-walled and $\sim$6 nm in diameter \cite{li.s:2012}.
During the fiber spinning, liquid infiltration like EG was used to densify the CNTs. By tuning the twisting speed, CNT
fibers with different twisting angles such as 10\si{\degree}, 17\si{\degree}, 23\si{\degree}, and 30\si{\degree} were
produced. The acid treatment was performed by immersing CNT fibers in concentrated \ce{HNO3} (16 M) for 2 h. After the
treatment, the CNT fibers were washed by water and dried under ambient conditions.

Epoxy resin (E44, fracture strain 10.5\%, Hangzhou Wuhuigang Adhesive Co., Ltd., Hangzhou, China), methyl
hexahydrophthalic anhydride (MHHPA, Puyang Huicheng Electronic Material Co., Ltd., Puyang, China), and imidazole curing
agent (Jiangxi Jinkai Chemical Co., Ltd., Jinxi, China) were mixed together to prepare the epoxy droplets, with a mass
ratio of 1:1:0.01. In some experiments the mixture was diluted with acetone, by adjusting the resin (including MHHPA and
imidazole) to acetone mass ratio to be 25:4 or 1:40. The curing profile was 90 \si{\celsius} for 1 h and 150
\si{\celsius} for 3 h.

The rheological measurements on the resin mixture were carried out in an oscillatory mode on a Bohlin Gemini 200
rheometer (Malvern Instruments, Worcestershire, United Kingdom). The tensile properties of CNT fiber was conducted with
the T150 Universal Testing Machine (Keysight Technologies, Inc., Santa Rosa, USA). The samples were mounted on paper
tabs with a gauge length of 7$\pm$1 mm, and the extension speed was set to 0.001 \si{\mm\per\s}. The microdroplet tests
were performed with the Equipment for Evaluation of Fiber/Resin Composite Interface Properties (model HM410, Tohei
Sangyo Co., Ltd., Tokyo, Japan). Figure \ref{fig.microdroplet} shows schematically the method of sample preparation. A
load cell of 1 N was used to characterize the debonding force. The drawing speed was 0.002 \si{\mm\per\s}.

\begin{acknowledgements}
The authors thank financial supports from the National Natural Science Foundation of China (11302241, 11404371,
21273269, 21473238, 51202251), Suzhou Industrial Science and Technology Program (ZXG201416), and the Youth Innovation
Promotion Association of the Chinese Academy of Sciences (2015256, Grant to X.Z.).
\end{acknowledgements}

%

\end{document}